\documentclass[lettersize,journal]{IEEEtran}

\usepackage{amssymb}
\usepackage{amsmath}
\usepackage{algorithmic}
\usepackage{algorithm}
\usepackage{array}
\usepackage{booktabs}
\usepackage{cases} 
\usepackage{caption}
\usepackage{comment}
\usepackage{cite}
\usepackage{caption}%
\usepackage{color}
\usepackage{footnote}
\usepackage{float}
\usepackage{graphicx}
\graphicspath{{images/}}
\DeclareGraphicsExtensions{.pdf,.jpeg,.png}
\usepackage{mathrsfs}
\usepackage{mathtools} 
\usepackage{multirow}
\usepackage{marvosym}
\usepackage[caption=false,font=normalsize,labelfont=sf,textfont=sf]{subfig}
\usepackage{silence}
\usepackage{tabularx}
\usepackage[para]{threeparttable}
\usepackage[colorinlistoftodos]{todonotes}
\usepackage[normalem]{ulem}
\useunder{\uline}{\ul}{}

\newcommand{\myshrink}[1]{%
  \ifnum#1>0
    \!\expandafter\myshrink\expandafter{\the\numexpr#1-1\relax}%
  \fi
}

\newcommand\reffigure[1]{Fig. \ref{#1}}

\newcommand\refproblem[1]{(\ref{#1})}
\newcommand\reftable[1]{Table \ref{#1}}

\makeatletter
\newcommand{\removelatexerror}{\let\@latex@error\@gobble}
\makeatother
\usepackage{array}

\usepackage[strings]{underscore}

\begin{document}
\title{$\ell_1$DecNet+: A new architecture framework by $\ell_1$ decomposition and iteration unfolding for sparse feature segmentation}

\author{Yumeng~Ren,~\IEEEmembership{~}
Yiming~Gao,~\IEEEmembership{~}
Xue-cheng~Tai,~\IEEEmembership{~}
and~Chunlin~Wu\textsuperscript{\Letter}~\IEEEmembership{~}
\thanks{Y. Ren is with the School of Mathematical Sciences, Nankai University, Tianjin, China and the Department of Mathematics, City University of Hong Kong, China (e-mail: ymren3-c@my.cityu.edu.hk).}
\thanks{Y. Gao is with the School of Mathematics, Nanjing University of Aeronautics and Astronautics, Nanjing, China (e-mail: gaoyiming@nuaa.edu.cn)}
\thanks{C. Wu is with the School of Mathematical Sciences, Nankai University, Tianjin, China (e-mail:  wucl@nankai.edu.cn ).}
\thanks{X. Tai is with the Norwegian Research Centre (NORCE), Bergen, Norway ( e-mail: xtai@norceresearch.no ).}
\thanks{Manuscript revised May 25, 2024.}}

\markboth{Journal of \LaTeX\ Class Files,~Vol.~14, No.~8, Feb~2024}%
{Shell \MakeLowercase{\textit{et al.}}: A Sample Article Using IEEEtran.cls for IEEE Journals}

\maketitle

\def\arraystretch{1.5} 

\begin{abstract}

     $\ell_1$ based sparse regularization plays a central role in compressive sensing and image processing. In this paper, we propose $\ell_1$DecNet, as an unfolded network derived from a variational decomposition model incorporating $\ell_1$ related sparse regularization and solved by scaled alternating direction method of multipliers (ADMM). $\ell_1$DecNet effectively decomposes an input image into a sparse feature and a learned dense feature, and thus helps the subsequent sparse feature related operations. Based on this, we develop $\ell_1$DecNet+, a learnable architecture framework consisting of our $\ell_1$DecNet and a segmentation module which operates over extracted sparse features instead of original images. This architecture combines well the benefits of mathematical modeling and data-driven approaches. To our best knowledge, this is the first study to incorporate mathematical image prior into feature extraction in segmentation network structures. Moreover, our $\ell_1$DecNet+ framework can be easily extended to 3D case. We evaluate the effectiveness of $\ell_1$DecNet+ on two commonly encountered sparse segmentation tasks: retinal vessel segmentation in medical image processing and pavement crack detection in industrial abnormality identification. Experimental results on different datasets demonstrate that, our $\ell_1$DecNet+ architecture with various lightweight segmentation modules can achieve equal or better performance than their enlarged versions respectively. This leads to especially practical advantages on resource-limited devices.
\end{abstract}



\begin{figure*}[!t]
    \centering
    \includegraphics[width=\textwidth]{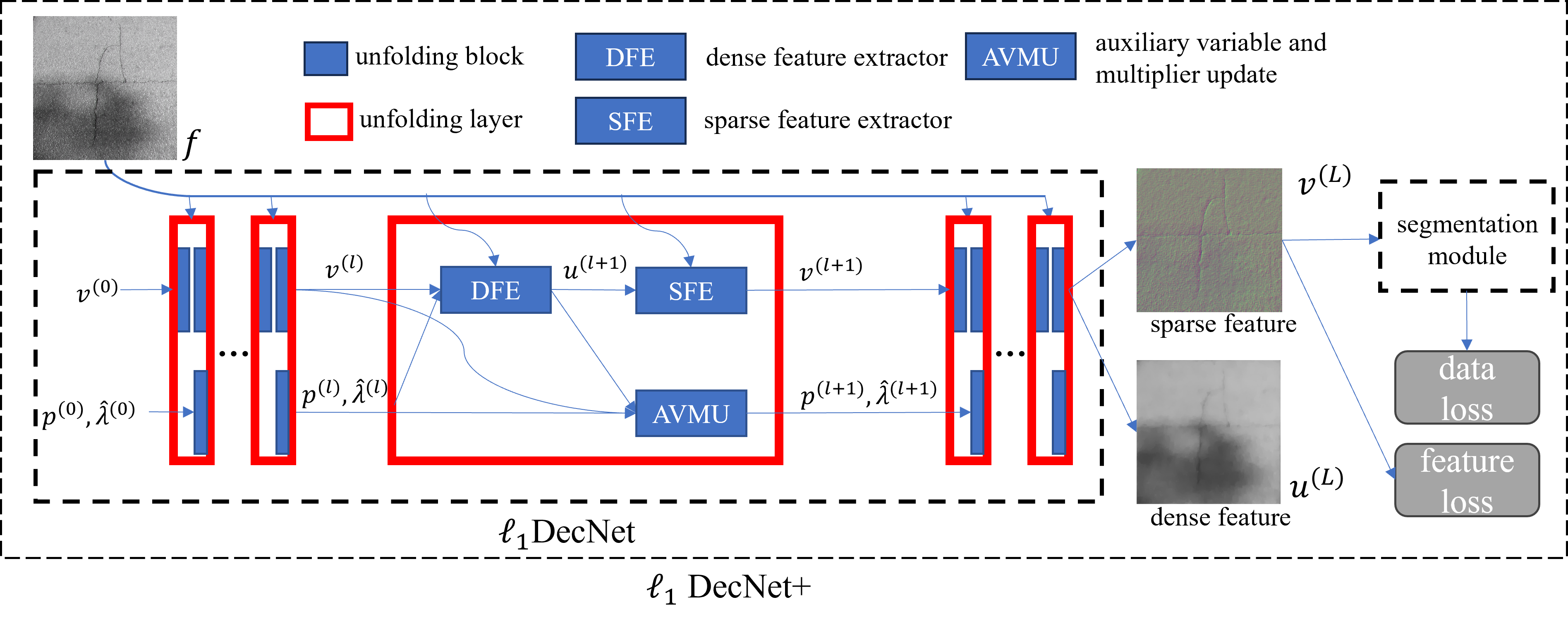}
    \caption{Our $\ell_1$DecNet+ architecture. The $\ell_1$DecNet with $L$ unfolding layers decomposes an input image $f$ into a sparse feature $v^{(L)}$ and a dense feature $u^{(L)}$, and the segmentation module operates over the sparse feature $v^{(L)}$ for sparse feature segmentation.}
    \label{fig:overview-idmunet}
\end{figure*}
\section{Introduction}

\IEEEPARstart{I}{mage} segmentation is one of the most important middle-level vision tasks, which bridges low-level vision operations and high-level vision applications.
So far, people developed lots of conventional methods and learning-based methods. In whatever method, an accurate suitable feature description of each object content is a key. In conventional segmentation methods, object features are explicitly modeled, such as a constant or polynomial intensity function. In learning-based methods, object features are implicitly learned in neural networks and usually lack of interpretability.

In this paper, we consider the problem of sparse feature extraction and segmentation from an input image. This problem raises in diverse applications like retinal vessel segmentation and crack detection. To solve this problem, we assume an input image be an addition of a sparse feature and a hard-to-describe dense feature (background). Our idea is to combine mathematical modeling and machine learning spirits, by introducing an $\ell_1$ decomposition model and using deep unfolding strategy.

Based on the powerful $\ell_1$ regularization, also even called ``modern least squares'', we propose a decomposition minimization model, which decomposes an input image to two components, one as the sparse feature and the other as the dense feature (background). The sparse feature component is characterized by an $\ell_1$ regularization, while the hard-to-describe background component is characterized by an $\ell_1$ regularization composed by some sparsifying linear transformations which will be learned from training data. After deriving a scaled-ADMM solver for this composite optimization problem, we unfold the iterative scheme to construct a deep neural network to build our $\ell_1$DecNet. Then an architecture framework named $\ell_1$DecNet+ is developed, by connecting our 
$\ell_1$DecNet and an any segmentation module; see \reffigure{fig:overview-idmunet}.$\ell_1$DecNet extracts a sparse feature well from an input image and deliver the feature to the subsequent segmentation module to finalize the segmentation task. As can be seen, sparsity priors (with or without learnable linear transformations) for feature descriptions are embedded into $\ell_1$DecNet+. Therefore, it combines well mathematical modeling and data-driven spirits. To train $\ell_1$DecNet+, we use ADAM to minimize a loss function consisting of a segmentation loss and a feature loss.

We conduct our experiments of $\ell_1$DecNet+ on three datasets DRIVE, CHASE\_DB1 and CRACK for sparse feature segmentation. We construct six $\ell_1$DecNet+ architectures by using six popular lightweight segmentation networks as the segmentation modules for our experiments. Tests and comparisons show that our $\ell_1$DecNet+ has the flexibility on the choice of segmentation module. While achieving an equal or better performance of the compared large network related to the used segmentation module, our $\ell_1$DecNet+ architecture uses much less learnable parameters and much less storage occupation for network weights.

Our contributions are as follows
\begin{itemize}

\item By modeling an input image as an addition of a sparse feature and a hard-to-describe dense feature (background), we propose an $\ell_1$ decomposition model, where the sparse feature is characterized by an $\ell_1$ regularization and the dense feature is characterized by an $\ell_1$ regularization composed by some sparsifying linear transformations.
\item Based on a scaled-ADMM solver for the $\ell_1$ decomposition model and deep unfolding method, we propose a learnable sparse feature extraction network $\ell_1$DecNet, where the linear sparsifying transformations for the hard-to-describe dense feature and some model and algorithm parameters are relaxed to be learnable.
\item Through delivering the extracted sparse feature by $\ell_1$DecNet to a segmentation module, we further construct an $\ell_1$DecNet+ architecture, which can adapt to diverse lightweight segmentation modules to achieve equal or better performances than their enlarged versions respectively, indicating the advantages of integrating mathematical modeling and data-driven strategy.
\end{itemize}

\section{Related Works}

In this section, we review most related works on (1)  $\ell_1$ regularization in compressive  image processing, (2) the variational framework for image decomposition with different regularization towards different tasks, and (3) deep unfolding methods.
\subsection{$\ell_1$ related regularization}

$\ell_1$ related regularization is one of the most powerful techniques to characterize sparsity property of signal and image data. It plays a central role in compressive sensing (CS) theory\cite{donoho_compressed_2006,candes_decoding_2005,candes_stable_2006} with extensive applications and was even called ``modern least squares"\cite{candes_enhancing_2008}, due to its computational tractability and capability to reconstruct sparse signals from fewer measurements than those required by Shannon-Nyquist sampling theorem. The $\ell_1$ regularization composed by certain linear transformations helps to design effective minimization models in imaging applications, like total variation (TV) model\cite{rudin_nonlinear_1992}, wavelet frame based approaches\cite{daubechies_iterative_2004,cai_convergence_2009,cai_image_2012}, and lots of subsequent developments (See, e.g., those summarized in \cite{scherzer_handbook_2010,chen_handbook_2023,liu2023variable}).

$\ell_1$ related regularization, although being convex, is non
smooth. A large variety of efforts were devoted to developing efficient optimization algorithms to solve $\ell_1 $ related minimization models, like iterative shrinkage-thresholding algorithm (ISTA)\cite{daubechies_iterative_2004}, Bregman iterations\cite{yin_bregman_2008, goldstein_split_2009}, fast iterative shrinkage-thresholding algorithm (FISTA)\cite{beck_fast_2009}, alternating direction method of multipliers (ADMM)\cite{wu_augmented_2010,boyd_distributed_2010}, primal-dual\cite{chambolle_first-order_2011,esser_general_2010}, proximity based fixed point iteration\cite{micchelli_proximity_2011};  see recent books\cite{glowinski_splitting_2017,chen_handbook_2023} and references therein.

We use $\ell_1$ related regularization with or without linear transformation to characterize the sparsity  of different feature components of an image.

\subsection{Variational image decomposition}

In image processing, one important task is to decompose one single image into two or more useful components. These components illustrate different image structures and benefit for many high-level vision tasks\cite{li_joint_2018,aujol_structure-texture_2006,kang_automatic_2011}.

In variational framework, the related minimization model should thus consist of two or more regularizers for different components and dates back to infimal convolution techniques\cite{chambolle_image_1997}. Following \cite{chambolle_image_1997}, people considered various decomposition models. Usually there is one cartoon component regularized by the TV prior\cite{chambolle_image_1997,setzer_infimal_2011} or non-convex TV prior\cite{duan_l-0_2015,chang_new_2017,guo_effective_2021}. The TV prior is a composition of $\ell_1$-norm and gradient operators, characterizing the sparsity \cite{chen_atomic_1998,donoho_compressed_2006,candes_robust_2006} of an image in gradient domain. The other components depend on applications, such as cartoon-texture decomposition\cite{jung_simultaneous_2015,ham_robust_2018,schaeffer_low_2013}, Retinex illumination estimation\cite{kimmel_variational_2003,ng_total_2011,ma_tv_2012,liang_retinex_2015} and intensity inhomogeneity removal\cite{chang_new_2017,guo_effective_2021} for segmentation. Note that these are not data-driven methods, and model and algorithm parameters therein are set manually.

\subsection{Deep unfolding methods}

From the pioneering work of LISTA (learned ISTA)\cite{gregor_learning_2010}, deep unfolding methods provide powerful tools for the network design on problems equipped with clear physical or mathematical models. The implicit layer-like structures in iterative solvers are mapped into deep architectures. Model and algorithm parameters, and even others, are relaxed to be learnable from data. 

As far as we know, there are two categories of deep unfolding methods. In the first type, people construct unfolding networks\cite{long_pde-net_2019,haber_stable_2017,chen_trainable_2017,chen_neural_2018,bai_deep_2019} from numerical ODE or PDE. In the second type, people unroll some optimization algorithms to construct networks, like \cite{gregor_learning_2010,wu_sparse_2020} from ISTA for sparse coding,  \cite{adler_learned_2018} from primal dual hybrid gradient algorithm for image reconstruction, \cite{zhang_ista-net_2018,song_memory-augmented_2021} from ISTA, \cite{yang_admm-csnet_2020} from ADMM and \cite{xiang_fista-net_2021} from FISTA for CS reconstruction, and \cite{zhang_amp-net_2021,fu_model-driven_2022,zhang_deep_2020,wu_uretinex-net_2022} from alternating minimization solving (approximate) penalized problems for image restoration or enhancement tasks; See \cite{chen_handbook_2023} and references therein for more details.

Unlike previous works, we start from an additive decomposition model and its ADMM solver, and design a learnable architecture for feature extraction and segmentation task.

\section{$\ell_1$DecNet from an $\ell_1$ decomposition model and its scaled-ADMM solver}

In this section, we build our $\ell_1$DecNet that decomposes an image into its sparse feature and dense feature components. We first introduce a variational decomposition model with $\ell_1$ related regularization. After presenting its scaled-ADMM solver, we construct an unfolding-based decomposition network, i.e., $\ell_1$DecNet.

\subsection{An $\ell_1$ variational decomposition model for sparse feature extraction}

We first give basic notations to construct our models. Without loss of generality, the 2D gray image $u$ can be represented as an $N \times N$ array. We denote $X=\mathbb{R}^{N \times N}$, and equip $X$ with inner product $\left\langle\cdot,\cdot\right\rangle$ and $\|\cdot \|$ as $\left\langle u,v \right\rangle =\sum_{i=1}^{N}\sum_{j=1}^{N}u_{i,j}v_{i,j}$, $\|u\|_2 = \sqrt{\left\langle u,u \right\rangle}$, $\|u\|_1 = \sum_{i=1}^{N}\sum_{j=1}^{N}|u_{i,j}|$, for $u,v \in X$; See, e.g., \cite{wu_augmented_2010,liu2023variable} for more related details.

For a gray-scale image $f \in X$, we introduce a variational model to decompose it into two features: a sparse feature $v \in X$, and a dense feature $u \in X$. According to \cite{chen_atomic_1998, donoho_compressed_2006, candes_robust_2006}, we use $\ell_1$ norm to characterize the sparsity of $v$. The dense feature $u$ is, however, assumed to be characterized by the sparsity in the domain of a group of convolution kernels. Our decomposition model is then given as follows:
\begin{align}
    \mathop{\min}\limits_{u,v\in X}
     \sum_{m=1}^{M}\alpha_m \|K_m u\|_1 +\beta \|v\|_1 + \frac{1}{2}\| u+v-f\|_2^2,
    \label{pr:model}
\end{align} where each $K_m: X \rightarrow X$ is a linear convolutional sparsifying transformation and $\alpha_m>0$, $\beta>0$ are some regularization parameters, with $m=1,2,\cdots,M$. We mention that an isotropic variant of model (\ref{pr:model}) was proposed in \cite{calatroni_infimal_2017} for mixed noise removal, in a non- data driven fashion. \cite{kim_structure-texture_2019} uses a similar constrained model for structure-texture decomposition, and embeds a pre-trained CAN\cite{yu_multi-scale_2016} network into their iterative algorithm, following the plug-and-play idea, instead of constructing an unrolling network structure suitable for end-to-end training.

\subsection{A scaled-ADMM iterative solver}\label{sec:idn-itersolver}

The model \refproblem{pr:model} is, although non-smooth, but convex and with separable structures. It can be efficiently solved by various splitting methods with convergence guarantee\cite{glowinski_splitting_2017}, such as ADMM, primal-dual methods and split Bregman iteration. In this paper, we take the scaled-ADMM \cite{glowinski_sur_1975,boyd_distributed_2010}. The algorithm uses variable splitting and the augmented Lagrangian function, which alternatively solves some simple subproblems.

We first introduce auxiliary variables $$p=(p_1,p_2,...,p_M) \in X^M = \underbrace{X \times X \cdots \times X}_{M}~,$$ and rewrite the model \refproblem{pr:model} as the following equivalent one \begin{align}
\left\{
\begin{aligned}
    \mathop{\min}\limits_{u,v \in X \atop p \in X^M} & \sum_{m=1}^{M} \alpha_m\|p_m\|_1 +\beta\|v\|_1 + \frac{1}{2}\|u+v-f\|_2^2,\\
    \text{s.t., } & p_m = K_m u,~m=1,2,\cdots,M.\\
\end{aligned}
\right.
\label{pr:constraint}
\end{align} We define its augmented Lagrangian function as 
\setlength{\arraycolsep}{0.0em} 
\begin{eqnarray}
    \thinmuskip=1mu
    \medmuskip=0mu
    \label{eq:al-function}
    &\mathscr{L}(u,v,p;\lambda)=& \frac{1}{2}\|u+v-f\|_2^2 + \beta\|v\|_1 \nonumber\\
    &{ }+\displaystyle\sum_{m=1}^{M} \bigg( \alpha_m\|p_m\|_1 &+\left\langle \lambda_m,K_mu-p_m\right\rangle+\frac{r}{2}\|K_mu-p_m\|_2^2 \bigg), \nonumber
\end{eqnarray} with Lagrange multipliers $\lambda=(\lambda_1, \lambda_2,\cdots,\lambda_M)$ and penalty parameter $r>0$, which is reformulated to be\begin{eqnarray}
    \thinmuskip=1mu
    \medmuskip=0mu
    \label{eq:al-scaled-function}
    &\widehat{\mathscr{L}}(u,v,p;\widehat{\lambda})=& \frac{1}{2}\|u+v-f\|_2^2 + \beta\|v\|_1  \nonumber \\
    &{ }+\displaystyle\sum_{m=1}^{M}\bigg( \alpha_m\|p_m\|_1 &+\frac{r}{2}\|K_mu-p_m+\widehat{\lambda}_m\|_2^2 - \frac{r}{2}\|\widehat{\lambda}_m\|^2_2\bigg)\nonumber
\end{eqnarray} by letting $\widehat{\lambda}_m = \lambda_m/r$, as in \cite{glowinski_sur_1975,boyd_distributed_2010}. We then amount to solve the saddle point problem below:
\begin{align}
    \begin{aligned}
    &\text{Find } (u^*,v^*,p^*;\widehat{\lambda}^*) \in X \times X  \times X^M \times X^M,\\
    &\text{s.t., } \widehat{\mathscr{L}}(u^*,v^*,p^*;\widehat{\lambda}) \le \widehat{\mathscr{L}}(u^*,v^*,p^*;\widehat{\lambda}^*) \le \widehat{\mathscr{L}}(u,v,p;\widehat{\lambda}^*),\\
    &\forall (u,v,p;\widehat{\lambda}) \in X \times X \times X^M \times X^M. 
    \end{aligned}
\label{pr:saddle}
\end{align} Its scaled-ADMM solver is presented in Algorithm~1:

\begin{algorithm}[H]
    \caption{Scaled-ADMM solver for (\protect\refeq{pr:saddle})}\label{alg:alg1}
    \begin{algorithmic}
    \STATE 
    \STATE 1.~{$\mathbf{Initialization}$} $p^{(0)},u^{(0)},v^{(0)},\widehat{\lambda}^{(0)} = 0$;
    \STATE 2.~{$\mathbf{For}$} $l = 0,1,\cdots$, until convergence:
    \STATE \hspace{0.5cm}Compute $u^{(l+1)}$ from
    \STATE \hspace{0.5cm}{\begin{align}
        \mathop{\min}\limits_{u\in X }~\widehat{\mathscr{L}}(u,v^{(l)},p^{(l)};\widehat{\lambda}^{(l)});
    \label{pr:admm-uv}
    \end{align}}
    \STATE \hspace{0.5cm}Compute $(v^{(l+1)},p^{(l+1)})$ from
    \STATE \hspace{0.5cm}{\begin{align}
        \mathop{\min}\limits_{(v,p)\in X\times X^M}&~\widehat{\mathscr{L}}(u^{(l+1)},v,p;\widehat{\lambda}^{(l)});
    \label{pr:admm-pq}
    \end{align}}
    \STATE \hspace{0.5cm}Update $\widehat{\lambda}^{(l+1)}$ by
    \STATE \hspace{0.5cm}{\begin{align} \begin{aligned}
        \widehat{\lambda}^{(l+1)}_m &= \widehat{\lambda}^{(l)}_m+K_mu^{(l+1)}-p^{(l+1)}_m,~1\le m\le M.
    \label{eq:admm-upd}
    \end{aligned}
    \end{align}}
    \end{algorithmic}
    \label{alg1}
\end{algorithm}

Note that each subproblem has an explicit solution. In \refproblem{pr:admm-uv}, computing $u^{(l+1)}$ is equivalent to solve a linear system:
\begin{equation}
    \thinmuskip=1mu
    \medmuskip=0mu
    (u + v^{(l)} - f ) + r \sum_{m=1}^{M}K_m^{\top}(K_m u-p_m^{(l)}+\hat{\lambda}_m^{(l)}) =0,
\end{equation}with explicit solution \begin{equation}
    \thinmuskip=1mu
    \medmuskip=0mu
    u^{(l+1)}= \left(\myshrink{4} \text{Id}+r\sum_{m=1}^{M}K_m^\top K_m\myshrink{4}\right)^{\myshrink{6}-1} \myshrink{6}\left(\myshrink{4}f-v^{(l)} + r\sum_{m=1}^{M}K_m^\top(p_m^{(l)} - \hat{\lambda}_m^{(l)})\myshrink{4}\right),
    \label{eq:solver-u}
\end{equation} where $\text{Id}$ represents the identity operator. For the convolutional operator $K_m$ with periodic boundary condition, \refproblem{pr:admm-uv} can be efficiently solved by FFT; see, e.g., \cite{wang_new_2008,wu_augmented_2010}.

As for \refproblem{pr:admm-pq}, we compute $(v^{(l+1)},p^{(l+1)})$ by separately solving the following:\begin{align}
    \left\{\begin{aligned}
    v^{(l+1)} &= \mathop{\arg\min}\limits_{v\in X }\left\{\beta\|v\|_1 + \frac{1}{2} \|u^{(l+1)}+v-f\|_2^2\right\}, \\
    p^{(l+1)}_m &= \mathop{\arg\min}\limits_{p_m\in X }\left\{\alpha_m \|p_m\|_1 + \frac{r}{2} \|K_m u^{(l+1)}-p_m + \hat{\lambda}^{(l)}_m\|_2^2\right\},
    \label{pr:admm-pv-para}
    \end{aligned}\right.
\end{align} where $~m=1,2,\cdots,M$. These parallelized problems have explicit solutions:\begin{align}
        v^{(l+1)}&=\mathscr{S}\left(f-u^{(l+1)};\beta\right),\label{eq:solver-v}\\
        p_m^{(l+1)}&=\mathscr{S}\left(K_mu^{(l+1)}+\hat{\lambda}_m^{(l)};\frac{\alpha_m}{r}\right),~1\le m \le M,\label{eq:solver-p}
\end{align} where $\mathscr{S}$ is the element-wise soft-thresholding operation \cite{donoho_denoising_1995,wu_augmented_2010} with \begin{align}
    \mathscr{S}\left(x;\gamma\right)=\left\{ 
    \begin{aligned}
        &\text{sgn}(x)(|x|-\gamma)\text{, }\\
        &0\text{, }
    \end{aligned}\right.
    \begin{aligned}
        &\text{ if }|x|\ge \gamma\text{, }\\
        &\text{ else.}
    \end{aligned}
\end{align} for $x \in \mathbb{R}$ and $\gamma \in \mathbb{R}_+$.

\begin{figure}
    \centering
    \includegraphics[width=0.4\textwidth]{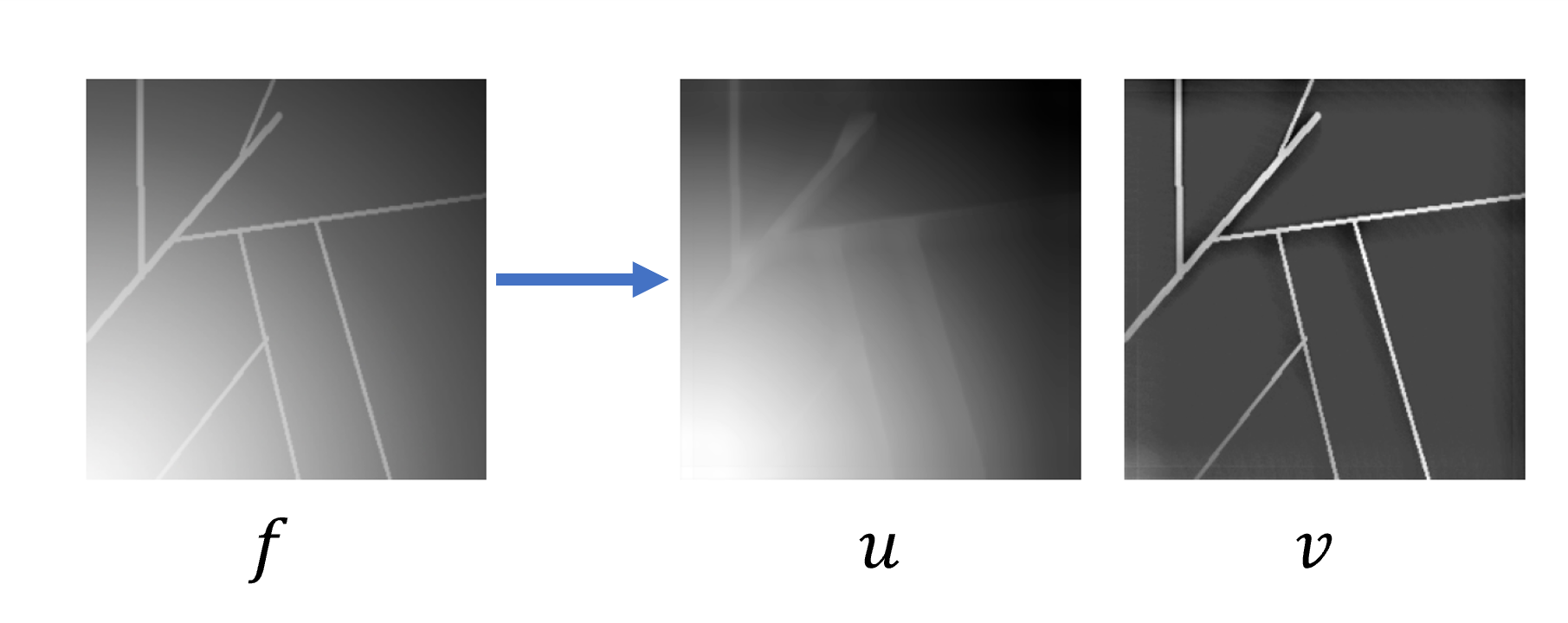}
    \caption{The decomposition result of the proposed model (\protect\refeq{pr:model}) with $M=2$, $(K_1,K_2)=(\nabla_x,\nabla_y)$, $\alpha_1=\alpha_2=0.006$, $\beta=0.003$ and using Algorithm~1 with $r=5$ on a synthetic image $f$.}
    \label{fig:id-fuv}
\end{figure}

With a proper selection of parameters, Algorithm 1 for $\ell_1$ decomposition model \refeq{pr:model} fulfils the purpose of sparse feature extraction. \reffigure{fig:id-fuv} shows an example of $\ell_1$ decomposition. 

Note that the above iterative algorithm is slightly different from the standard scaled-ADMM. We do not introduce an auxiliary variable for $v$, to save the computational costs of the algorithm and the later unfolded network.

\subsection{$\ell_1$DecNet for sparse feature extraction}\label{sec:decnetwork}

In this subsection, our $\ell_1$DecNet, a trainable feature extractor, is constructed from Algorithm 1, following the idea of deep unfolding\cite{gregor_learning_2010,yang_deep_2016,adler_learned_2018,yang_admm-csnet_2020}. We mention that these existing literatures all start from reconstruction models and their constructed networks are for image restoration problems. Here our $\ell_1$DecNet is designed based on a variational decomposition model with two feature layers as outputs. We fix the first $L$ iterations and map each iteration into one network layer. The subproblem solvers are generalized to network blocks by relaxing the parameters and sparsifying transformations to be learnable, yielding the dense feature extractor (DFE) block, the sparse feature extractor (SFE) block and auxiliary variable and multiplier update (AVMU) block.

Our $\ell_1$DecNet is shown in \reffigure{fig:overview-idmunet}. The network input is $f$. We set the initial values $(v^{(0)}, p^{(0)},\widehat{\lambda}^{(0)})= (0,0,0)$ in the first layer. The network state for $l$-th layer ($l=0,1,\cdots,L-1$) is defined as $( v^{(l+1)}, p^{(l+1)},\widehat{\lambda}^{(l+1)})$, and $u^{(l+1)}$ is not passed to the next layer. The DFE, SFE and AVMU blocks are detailed as follows.

\subsubsection{DFE block}

The DFE block is induced from (\refeq{eq:solver-u}), which outputs the dense feature component $u^{(l+1)}$. Given $v^{(l)}$, $p^{(l)}$ and $\widehat{\lambda}^{(l)}$, the output of $l$-th $\text{DFE}$ is as follows 
\begin{multline}
    \thinmuskip=1mu
    \medmuskip=0mu
    u^{(l+1)}={}\\
    {}\left(\text{Id}+r\sum_{m=1}^M{K^{(l)}_m}^{\top}K^{(l)}_m\right)^{\myshrink{6}-1}\myshrink{3}\left(f-v^{(l)}+\sum_{m=1}^M r{K_m^{(l)}}^{\top}(p_m^{(l)}-\hat{\lambda}_m^{(l)})\right),
\end{multline} where $r$ and $K_m^{(l)}$ are trainable parameters. Note that we generalize $K_m$ in Algorithm~1 to be $K_m^{(l)}$, in order to increase the network capacity.

\subsubsection{SFE block}

The SFE block is induced from (\refeq{eq:solver-v}), which outputs the sparse feature component $v^{(l+1)}$. Given $u^{(l+1)}$, the output of $l$-th $\text{SFE}$ block is as follows \begin{align}
    v^{(l+1)}&=\mathscr{S}\left(f-u^{(l+1)};\beta^{(l)}\right),
\end{align} where $\beta^{(l)}$ is trainable parameter generalized from $\beta$ .  

\subsubsection{AVMU block}

The AVMU block is from (\refeq{eq:solver-p})(\refeq{eq:admm-upd}), which updates the auxiliary variables $p_m^{(l)}$ and multipliers  $\hat{\lambda}_m^{(l)}$ according to $u^{(l+1)}$, respectively.  Given $u^{(l+1)}$ and $(p^{(l)},\widehat{\lambda}^{(l)})$, the computational sequence is defined as \begin{align}
    \left\{
        \begin{aligned}
        p_m^{(l+1)}  &= \mathscr{S} (K_m^{(l)} u^{(l+1)}+\widehat{\lambda}_m^{(l)}; \hat\alpha_m^{(l)}),\\
        \widehat{\lambda}_m^{(l+1)} &= \widehat{\lambda}_m^{(l)} + K_m^{(l)} u^{(l+1)} -p_m^{(l+1)},~1 \le m \le M.
        \end{aligned}
    \right.
\end{align} where $K_m^{(l)}$ and $\hat \alpha_m^{(l)}$ are trainable. Note that $\hat\alpha_m^{(l)}$ replaces $\frac{\alpha_m^{(l)}}{r}$ to avoid zero division in the training procedure.

We now clarify the hyperparameters and learnable parameters of $\ell_1$DecNet. In $\ell_1$DecNet, we have the following hyperparameters, the number of layers $L$, the number of convolution kernels $M$, and the sizes of convolution kernels which are in this paper all set as $(2R+1)\times(2R+1)$ with an integer $R\ge0$. The learnable parameters include the number $r$ in DFE block shared by all layers, convolutional kernels $K_m^{(l)}$ in all three blocks of each layer, $\beta^{(l)}$ in SFE block and $\hat\alpha_m^{(l)}$ in AVMU block of each layer, where $m\in \{1,2,\cdots,M\}$ and $l\in\{0,1,\cdots, L-1\}$.

Calculations in $\ell_1$DecNet are all basic, including element-wise non-linear functions and 2D convolutions. Therefore, the back-propagation can be conducted efficiently by the Autograd package of deep learning framework like PyTorch. Also, it is straightforward to extend $\ell_1$DecNet to 3D case by using 3D convolutions.

Since our $\ell_1$DecNet is induced from the mathematical decomposition model (\refeq{pr:model}) and its convergent scaled-ADMM solver, the network computational procedure has good underlying mechanism and thus it actually does not required to learn too many parameters, as will be shown in our experiments.


\section{$\ell_1$DecNet+ for sparse feature segmentation}\label{sec:l1dplus}

We now present our $\ell_1$DecNet+ architecture for sparse feature segmentation, which performs segmentation operation on the sparse feature extracted by $\ell_1$DecNet. \reffigure{fig:overview-idmunet} shows the overall structure of $\ell_1$DecNet+, $\ell_1$DecNet followed by a segmentation module. $\ell_1$DecNet decomposes an input image into a dense feature $u$ and a sparse feature $v$. The following segmentation module takes only the feature $v$ for segmentation. This segmentation module is rather abstract and any current segmentation networks (like UNet and its variants) can be used here. We mention that, for multi-channel input images, each channel is processed by a separate $\ell_1$DecNet, and the extracted $v$ features of all channels are concatenated and delivered to the segmentation module.

We next give the training method for our $\ell_1$DecNet+ architecture. Given data pairs $(F,T) =\{(f^i,t^i)\}_{i=1}^D$ with images and labels, we define the training loss as

\begin{align}
    \begin{multlined}
    \text{Loss}(F,T;\theta_1,\theta_2) := \\\frac{1}{D}\sum_{i=1}^{D}\bigg[\underbrace{\|s^i_{\theta_1,\theta_2}(f^i)-t^i\|_2^2}_{\text{loss}_{\text{data}}}  +c\underbrace{\|v^i_{\theta_1}(f^i)\|_1}_{\text{loss}_{\text{feature}}}\bigg],
    \end{multlined}
\end{align} where \begin{align}
    (u^i_{\theta_1}(f^i),v^i_{\theta_1}(f^i)) &:= \mathcal{A}_{\theta_1}(f^i),\\
    s^i_{\theta_1,\theta_2}(f^i) &:= \mathcal{B}_{\theta_2}(v^i_{\theta_1}(f^i)).
\end{align} Here $u^i_{\theta_1}$, $v^i_{\theta_1}$ and $s^i_{\theta_1,\theta_2}$ are the dense feature, sparse feature and segmentation result of the $i$-th input $f^i$. The $\theta_1,\theta_2$ represent learnable parameters in $\ell_1$DecNet $\mathcal{A}_{\theta_1}$ and segmentation module $\mathcal{B}_{\theta_2}$, and $c$ is a positive coefficient. 
The feature loss aims to pursuit the sparsity assumption on $v$ through training. It prevents  $\ell_1$DecNet from degenerating to identity or all the kernels $K_m$ from degenerating to identity, thus avoids yielding $v \approx f$ or $v \approx u$. Such loss function will be optimized by ADAM algorithm.

The $\ell_1$DecNet+ architecture and training method combine well the mathematical modeling and data-driven spirit. It leads to a feature-aware segmentation method. With the help of $\ell_1$ decomposition for sparse feature extraction, it can be expected that our proposed method has better performance and can reduce the amount of trainable parameters. As far as we know, our $\ell_1$DecNet+ is the first one to integrate $\ell_1$ decomposition, the mathematical modeling on sparsity, into design of network blocks.

\section{Experiments}\label{sec:expr}

\begin{figure}[t]  
    \flushleft
    \subfloat[DRIVE]
    {\includegraphics[width=0.23\textwidth]{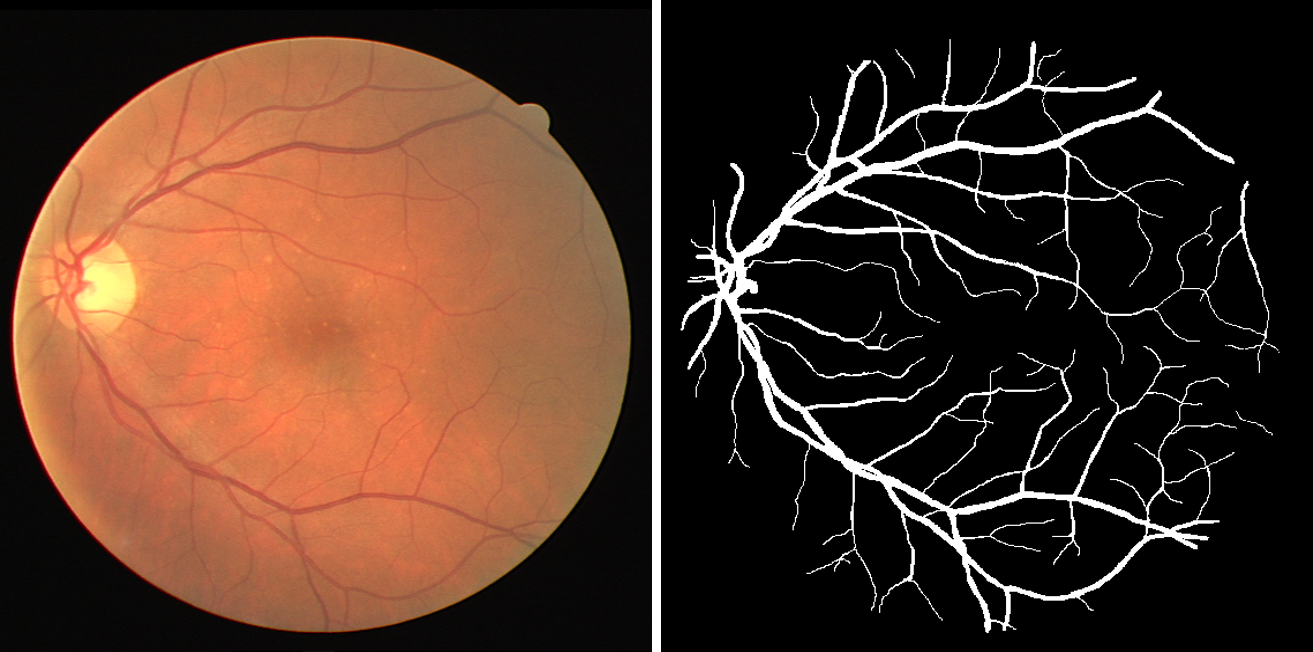}~}%
    \subfloat[CHASE]
    {\includegraphics[width=0.23\textwidth]{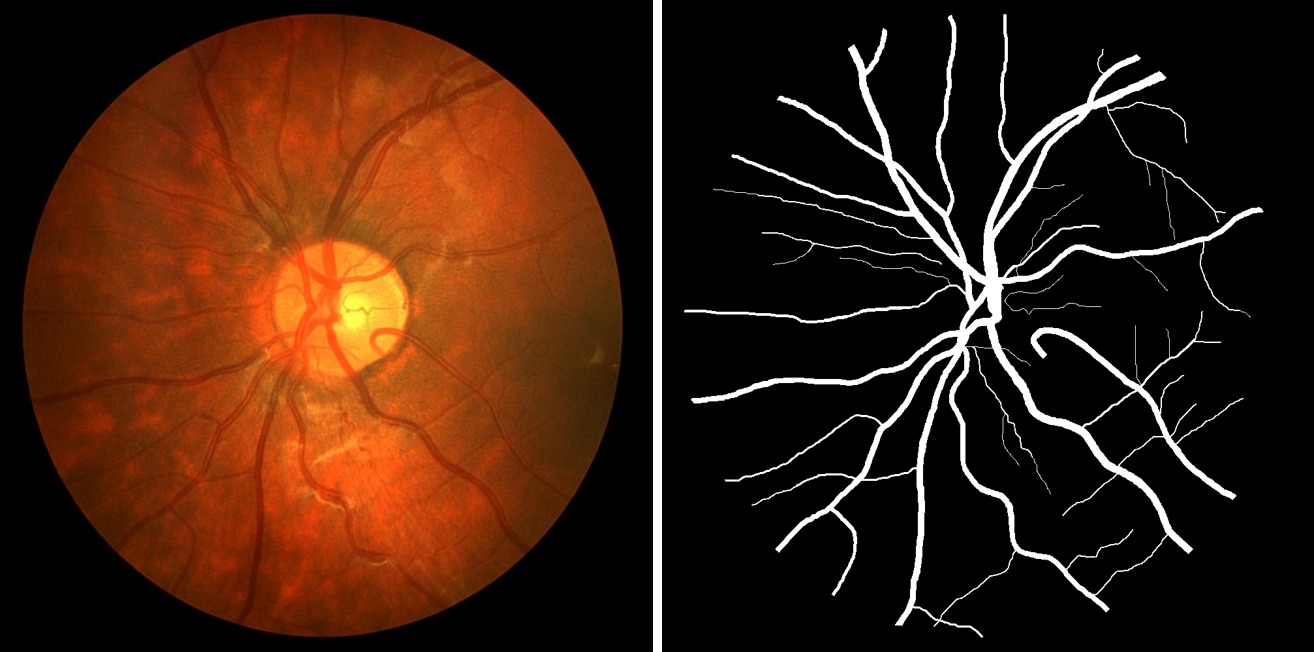}}%
    \hfill
    \flushleft
    \subfloat[CRACK]{\includegraphics[width=0.23\textwidth]{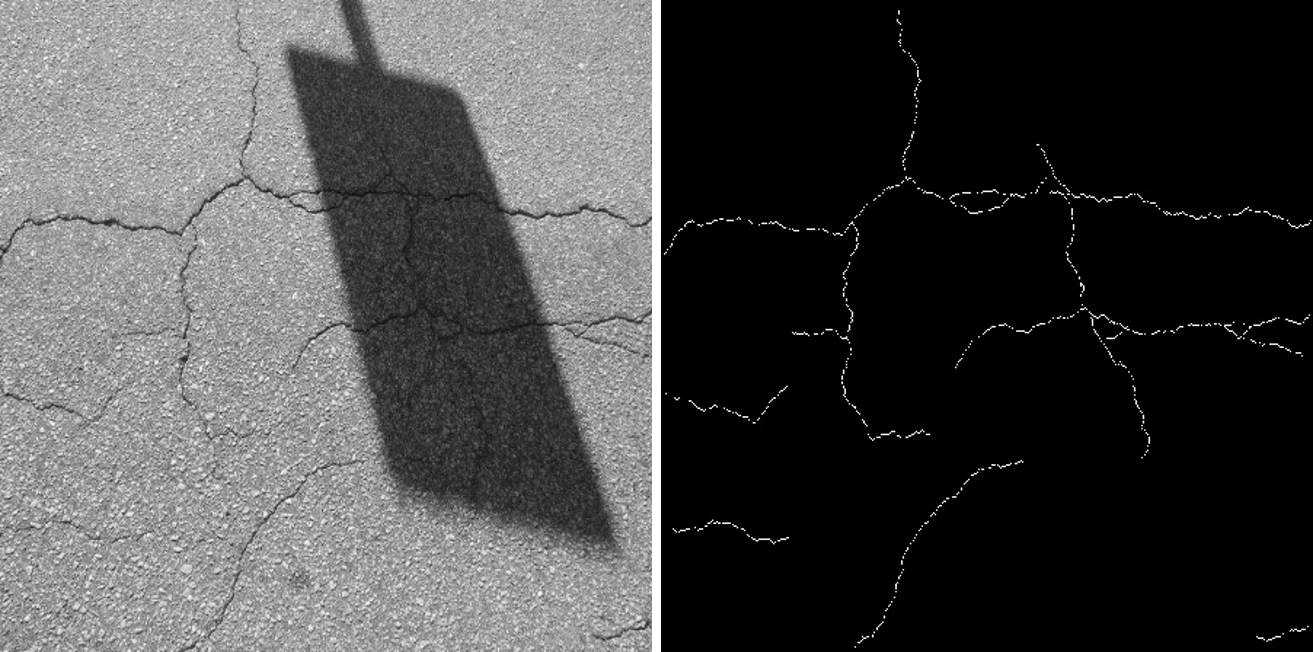}~}
    \subfloat[zoom-in of (c)]{\includegraphics[width=0.23\textwidth]{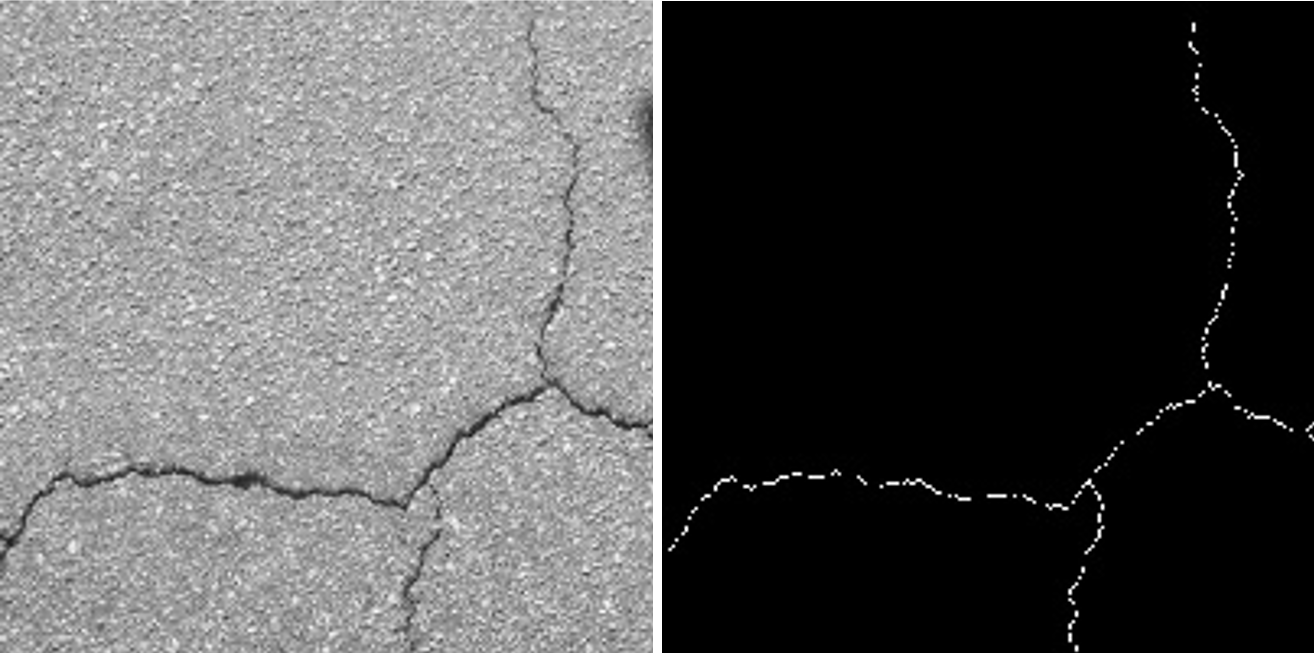}}
    \caption{Example images and their labels from DRIVE (a), CHASE (b) and CRACK (c)(d) datasets.}
    \label{fig:dataset}
\end{figure}

In this section, we show our experiments on our $\ell_1$DecNet+ architecture and comparisons to others for sparse feature segmentation. We implement $\ell_1$DecNet+ networks, training and testing procedures with PyTorch 1.12 and PyTorchLightning 1.7, using the ADAM optimizer and ReduceLROnPlateau learning rate scheduler. The segmentation modules and compared networks are taken from SegmentationMethodPytorch (SMP)\cite{Iakubovskii:2019} without pre-trained weights, and embedded into our training and testing procedures. All experiments are conducted on one GPU device, RTX 4090 24G, on a Ubuntu20.04 server.

\subsection{Datasets}\label{sec:expr-datapre}

Due to limited computational resources, we choose or construct the following small datasets for our experiments, i.e., DRIVE, CHASE (CHASE\_DB1) and CRACK.

\textbf{DRIVE}\cite{staal_ridge-based_2004} is a retinal vessel dataset with 40 RGB 565 $\times$ 584 images, and binary labels for 8.63\% vessel pixels. We take 20 images for training and the other 20 images for testing.

\textbf{CHASE} \cite{fraz_ensemble_2012} is a retinal vessel dataset with 28 RGB 999 $\times$ 960 images, and binary labels for 7.19\% vessel pixels. We take 20 images for training and 8 images for testing. 

\textbf{CRACK} is a pavement crack dataset with 206 RGB $448 \times 448$ images, and binary labels for 0.32\% cracks pixels. It consists of crack-centered cropped images of a subset of \textbf{Cracktree}\cite{zou_cracktree_2012}. We take 190 images for training and 16 images for testing.

\reffigure{fig:dataset} shows some images and labels from the above three datasets. During the training process, all images are randomly cropped into patches of a uniform size $64\times64$. During the testing process, images will be processed with overlapped moving window of $64\times64$. We use the standard flipping and cropping data augmentation from PyTorch.

\subsection{Choice of the segmentation module in $\ell_1$DecNet+}\label{sec:segbase}

The segmentation module (the segmentation subnetwork) of $\ell_1$DecNet+ in our experiments uses a UNet\cite{ronneberger_u-net_2015} of a certain size and with a certain single encoder, or a MANet\cite{fan2020ma}/UNet++ \cite{zhou_unet_2020} of a certain size and with fused multiple encoders. Four families of encoding blocks for UNet will be tested, denoted as CNN, Res-x, Eff-x and Mit-x, corresponding to UNet (2015)\cite{ronneberger_u-net_2015}, ResNet (2016)\cite{he_deep_2016}, EfficientNet (2019) \cite{tan_efficientnet_2019} and MixVisionTransformer (2021) \cite{xie2021segformer}. Besides, one family of encoding blocks for MANet/UNet++ is further tested, denoted as Res-x, corresponding to ResNet. The suffix x indicates the size of the encoding block. Documents of the github repository \cite{Iakubovskii:2019} show details of the structure of the segmentation module. For convenience of description, we abbreviate UNet, MANet and UNet++ as U, MA and Upp. We use subscripts to indicate the size of a segmentation module, and parentheses to indicate the encoding block. For example, $\text{U}_{3,32}$(Res18) represents a UNet of 3 down-samplings and 32 start channels with the ResNet18 encoding block. In particular, we test the following six small-scale segmentation modules in our $\ell_1$DecNet+ architecture, including $\text{U}_{2,16}$(CNN), $\text{U}_{3,32}$(Res18), $\text{U}_{3,32}$(Effb0), $\text{U}_{3,32}$(Mitb0), $\text{MA}_{3,32}$(Res18) and $\text{Upp}_{3,32}$(Res18).

\subsection{Hyperparameter settings and learnable parameter initializations }\label{sec:init-param}

Unless otherwise specified, we set hyperparameters and initialize learnable parameters in $\ell_1$DecNet+ and its training procedure as follows. In $\ell_1$DecNet, we set $L=8$, $M=2$ and $R=3$. The learnable parameters are initialized as in \reftable{tab:init-param}, where  $\nabla_x$ is a $(2R+1) \times (2R+1)$ kernel by zero-padding $\left(\begin{smallmatrix}
0&~1&~0\\
0&~ -\myshrink{2}1&~0\\
0&~0&~0\end{smallmatrix}\right)$and $\nabla_y$ is a $(2R+1) \times (2R+1)$ kernel by zero-padding $\left(\begin{smallmatrix}
0&~0&~0\\
0&~1&~-\myshrink{2}1\\
0&~0&~0\end{smallmatrix}\right)$. For the segmentation module in $\ell_1$DecNet+ and its enlarged version, we take the default parameter settings and initializations in the literatures. In the training loss, we set $c=0.1$.

\begin{table}[!h]
    \centering
    \caption{Initialization of learnable parameters in $\ell_1$DecNet for different datasets.}
        \begin{tabular}{l|llll}
        \hline
        \textbf{Dataset} & $\{K^{(l)}_m\}_{m=1}^M$   & $\hat\alpha_m^{(l)}$ & $\beta^{(l)}$ & $r$ \\ \hline
        \textbf{DRIVE/CHASE}  & $\{\nabla_x,\nabla_y\}$ & 0.006             & 0.003       & 5 \\ 
        \textbf{CRACK}  & $\{\nabla_x,\nabla_y\}$ & 0.006             & 0.02       & 5 \\ \hline
        \end{tabular}
    \label{tab:init-param}
\end{table}

\subsection{Evaluation metrics}
We use AUC (area under the ROC curve) score \cite{goodfellow2016deep} to evaluate the network performances in all experiments. The AUC score is a number in $[0,1]$, and is the higher the better. It is one of the most popular scores especially for binary segmentation and classification tasks in deep learning. 

Considering the randomness involved in the training process, these scores are calculated by averaging the scores of all the inferences obtained from 5 independent training and testing procedures of the network under the same settings.

\subsection{Experiments on the influence of hyperparameters of $\ell_1$DecNet}\label{sec:expr-alter}

Here we test the influence of hyperparameters $L$, $M$ and $R$ of $\ell_1$DecNet to the performance of $\ell_1$DecNet+ architecture, where we use $\text{U}_{2,16}$(CNN) as the segmentation module and DRIVE as the dataset. The AUC scores is shown in \reftable{tab:alt-mr}. We can see that, when $L=8$ is fixed, $M$ and $R$ have little influence on AUC score; meanwhile, when $M=2$ and $R=3$ are fixed, the AUC score slightly increases and peaks at $L=8$. Therefore in the following experiments, we use $L=8$ and $R=3$. As for the kernel number, we set $M=2$, considering the trade-off of the performance and memory occupation. 
\begin{table}[]
  \centering
    \caption{Experiments on the influence of different combinations of hyperparameters  $M$, $R$ and $L$ in terms of AUC for $\ell_1$DecNet+$\text{U}_{2,16}$(CNN) on DRIVE.}
    \label{tab:alt-mr}
    \begin{tabular}{l|ccc}
      \hline
          & \multicolumn{3}{c}{$L$=8,$R$=3}                           \\ \hline
          & $M$=2             & $M$=4             & $M$=6             \\ \hline
      AUC & 0.9862$\pm$0.0035 & 0.9867$\pm$0.0024 & 0.9863$\pm$0.0056 \\ \hline
          & \multicolumn{3}{c}{$L$=8,$M$=2}                           \\ \hline
          & $R$=2             & $R$=3             & $R$=4             \\ \hline
      AUC & 0.9857$\pm$0.0049 & 0.9862$\pm$0.0035 & 0.9861$\pm$0.0033 \\ \hline
          & \multicolumn{3}{c}{$M$=2, $R$=3}                          \\ \hline
          & $L$=6             & $L$=8             & $L$=10            \\ \hline
      AUC & 0.9859$\pm$0.0056 & 0.9862$\pm$0.0035 & 0.9857$\pm$0.0032 \\ \hline
      \end{tabular}
    \end{table}

\subsection{Experiments on the benefits by $\ell_1$DecNet feature extraction for $\ell_1$DecNet+ segmentation}\label{lab:result-3x}

\begin{table*}[]
  \centering
  \caption{Inference results (AUC) by different segmentation networks on three datasets. The $\ell_1$DecNet+ architecture with a small-scale segmentation module performs better than other segmentation networks in most cases,  while introducing few extra learnable parameters over its segmentation module.The symbol ``-'' means that the model has not generated reasonable results under current settings, and hence no score is reported; See \protect\reffigure{fig:train-seg-2}.}
  \label{tab:infer-3x}
  \begin{tabular}{l|ccc|ccc}
    \hline
     &
      $\text{U}_{3,32}$(CNN) &
      $\text{U}_{2,16}$(CNN) &
      \begin{tabular}[c]{@{}c@{}}$\ell_1$DecNet+\\ $\text{U}_{2,16}$(CNN)\end{tabular} &
      $\text{U}_{3,32}$(Res34) &
      $\text{U}_{3,32}$(Res18) &
      \begin{tabular}[c]{@{}c@{}}$\ell_1$DecNet+\\ $\text{U}_{3,32}$(Res18)\end{tabular} \\ \hline
    \#Param.   & 2.9m                    & 176k              & 2.5k+176k               & 21.5m                   & 11.4m             & 2.5k+11.4m              \\ \hline
    DRIVE(AUC) & 0.9722$\pm$0.0122       & 0.9741$\pm$0.0069 & {\ul 0.9862$\pm$0.0035} & 0.9674$\pm$0.0094       & 0.9680$\pm$0.0088 & {\ul 0.9851$\pm$0.0040} \\ \cline{1-1}
    CHASE(AUC) & 0.9744$\pm$0.0073       & 0.9719$\pm$0.0061 & {\ul 0.9847$\pm$0.0056} & 0.9854$\pm$0.0038       & 0.9845$\pm$0.0034 & {\ul 0.9892$\pm$0.0024} \\ \cline{1-1}
    CRACK(AUC) & {\ul 0.9905$\pm$0.0101} & 0.9859$\pm$0.0138 & 0.9874$\pm$0.0139       & {\ul 0.9908$\pm$0.0088} & 0.9893$\pm$0.0108 & 0.9905$\pm$0.0104       \\ \hline
     &
      $\text{U}_{3,32}$(Effb2) &
      $\text{U}_{3,32}$(Effb0) &
      \begin{tabular}[c]{@{}c@{}}$\ell_1$DecNet+\\ $\text{U}_{3,32}$(Effb0)\end{tabular} &
      $\text{U}_{3,32}$(Mitb1) &
      $\text{U}_{3,32}$(Mitb0) &
      \begin{tabular}[c]{@{}c@{}}$\ell_1$DecNet+\\ $\text{U}_{3,32}$(Mitb0)\end{tabular} \\ \hline
    \#Param.   & 7.82m                   & 4.1m              & 2.5k+4.1m               & 11.4m                   & 3.4m              & 2.5k+3.4m               \\ \hline
    DRIVE(AUC) & 0.9756$\pm$0.0074       & 0.9772$\pm$0.0074 & {\ul 0.9861$\pm$0.0039} & 0.9802$\pm$0.0067       & 0.9806$\pm$0.0054 & {\ul 0.9832$\pm$0.0045} \\ \cline{1-1}
    CHASE(AUC) & 0.9844$\pm$0.0045       & 0.9860$\pm$0.0030 & {\ul 0.9885$\pm$0.0038} & 0.9847$\pm$0.0034       & 0.9829$\pm$0.0040 & {\ul 0.9885$\pm$0.0024} \\ \cline{1-1}
    CRACK(AUC) & 0.9904$\pm$0.0116       & 0.9895$\pm$0.0102 & {\ul 0.9907$\pm$0.0137} & - & 0.9895$\pm$0.0113 & {\ul0.9900$\pm$0.0111}       \\ \hline
     &
      $\text{MA}_{3,32}$(Res34) &
      $\text{MA}_{3,32}$(Res18) &
      \begin{tabular}[c]{@{}c@{}}$\ell_1$DecNet+\\ $\text{MA}_{3,32}$(Res18)\end{tabular} &
      $\text{Upp}_{3,32}$(Res34) &
      $\text{Upp}_{3,32}$(Res18) &
      \begin{tabular}[c]{@{}c@{}}$\ell_1$DecNet+\\ $\text{Upp}_{3,32}$(Res18)\end{tabular} \\ \hline
    \#Param.   & 21.8m                   & 11.7m             & 2.5k+11.7m              & 21.5m                   & 11.4m             & 2.5k+11.4m              \\ \hline
    DRIVE(AUC) & 0.9733$\pm$0.0075       & 0.9746$\pm$0.0075 & {\ul 0.9851$\pm$0.0040} & 0.9739$\pm$0.0071       & 0.9681$\pm$0.0096 & {\ul 0.9860$\pm$0.0035} \\ \cline{1-1}
    CHASE(AUC) & 0.9855$\pm$0.0032       & 0.9841$\pm$0.0028 & {\ul 0.9872$\pm$0.0032} & 0.9863$\pm$0.0033       & 0.9847$\pm$0.0029 & {\ul 0.9893$\pm$0.0029} \\ \cline{1-1}
    CRACK(AUC) & 0.9879$\pm$0.0119       & 0.9876$\pm$0.0153 & {\ul 0.9885$\pm$0.0148} & {\ul 0.9900$\pm$0.0128} & 0.9879$\pm$0.0116 & 0.9887$\pm$0.0122       \\ \hline
    \end{tabular}
\end{table*}

\begin{table*}[]
  \centering
  \caption{Training, inference and storage costs in terms of time per backpropagation step (time per bp step), forward MACs (fwd MACs) and disk occupation, of different segmentation networks tested on CRACK. In general, the $\ell_1$DecNet+ architecture slightly increases the occupation in space and time over its segmentation module.}
\label{tab:train-3x}
\begin{tabular}{l|ccc|ccc}
  \hline
   &
    $\text{U}_{3,32}$(CNN) &
    $\text{U}_{2,16}$(CNN) &
    \begin{tabular}[c]{@{}c@{}}$\ell_1$DecNet+\\ $\text{U}_{2,16}$(CNN)\end{tabular} &
    $\text{Res34}_{3,32}$(Res34) &
    $\text{U}_{3,32}$(Res18) &
    \begin{tabular}[c]{@{}c@{}}$\ell_1$DecNet+\\ $\text{U}_{3,32}$(Res18)\end{tabular} \\ \hline
  time per bp step & 1.879 s    & 0.937 s   & 0.980 s   & 0.9408 s  & 0.9138 s  & 1.528 s   \\ \cline{1-1}
  fwd MACs         & 276.82 G   & 47.82 G   & 55.21 G   & 96.54 G   & 74.79 G   & 82.19 G   \\ \cline{1-1}
  disk occupation  & 34,323 KB  & 2,161 KB  & 2,266 KB  & 94,874 KB & 50,118 KB & 50,224 KB \\ \hline
   &
    $\text{U}_{3,32}$(Effb2) &
    $\text{U}_{3,32}$(Effb0) &
    \begin{tabular}[c]{@{}c@{}}$\ell_1$DecNet+\\ $\text{U}_{3,32}$(Effb0)\end{tabular} &
    $\text{U}_{3,32}$(Mitb1) &
    $\text{U}_{3,32}$(Mitb0) &
    \begin{tabular}[c]{@{}c@{}}$\ell_1$DecNet+\\ $\text{U}_{3,32}$(Mitb0)\end{tabular} \\ \hline
  time per bp step & 1.227 s    & 0.994 s   & 1.094 s   & 1.109 s   & 0.935 s   & 0.992 s   \\ \cline{1-1}
  fwd MACs         & 38.95 G    & 36.41 G   & 43.81 G   & 62.01 G   & 22.15 G   & 29.55 G   \\ \cline{1-1}
  disk occupation  & 32,216 KB  & 17,049 KB & 17,154 KB & 65,282 KB & 16,971 KB & 17,037 KB \\ \hline
   &
    $\text{MA}_{3,32}$(Res34) &
    $\text{MA}_{3,32}$(Res18) &
    \begin{tabular}[c]{@{}c@{}}$\ell_1$DecNet+\\ $\text{MA}_{3,32}$(Res18)\end{tabular} &
    $\text{Upp}_{3,32}$(Res34) &
    $\text{Upp}_{3,32}$(Res18) &
    \begin{tabular}[c]{@{}c@{}}$\ell_1$DecNet+\\ $\text{Upp}_{3,32}$(Res18)\end{tabular} \\ \hline
  time per bp step & 1.037 s    & 1.015 s   & 2.428 s   & 1.497 s   & 0.9828 s  & 1.868 s   \\ \cline{1-1}
  fwd MACs         & 81.22 G    & 59.48 G   & 66.88 G   & 115.45 G  & 93.70 G   & 101.10 G  \\ \cline{1-1}
  disk occupation  & 100,404 KB & 55,648 KB & 55,710 KB & 96,292 KB & 51,536 KB & 51,598 KB \\ \hline
  \end{tabular}
\end{table*}

\begin{figure}
  \centering 
  \includegraphics[width=0.5\textwidth]{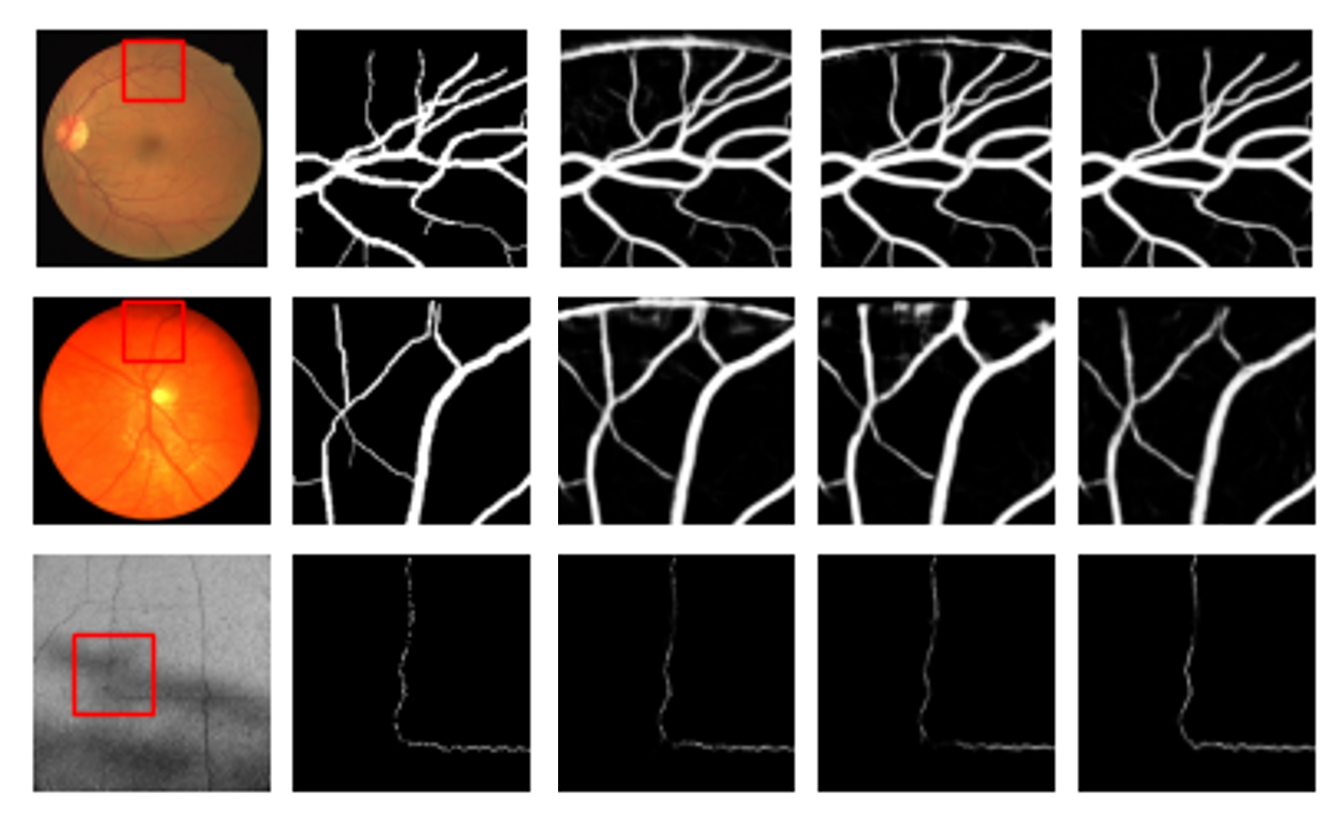}
  \caption{The input $f$ (column 1), zoom-in label (column 2) and zoom-in segmentation results from $\text{U}_{3,32}$(CNN), $\text{U}_{2,16}$(CNN) and $\ell_1$DecNet+$\text{U}_{2,16}$(CNN) (column 3-5). Images and labels are chosen from the test subset of DRIVE, CHASE and CRACK (row 1-3).} 
  \label{fig:seg}
\end{figure}

In this subsection, we show the benefits of $\ell_1$DecNet feature extraction for segmentation in $\ell_1$DecNet+ architecture, including the flexibility of the choice of segmentation module and the assistance on reducing the scale of the segmentation network.

We construct six $\ell_1$DecNet+ segmentation architectures by using small-scale segmentation modules, and compare them to their segmentation modules and enlarged versions. In particular, we compare $\ell_1$DecNet+ $\text{U}_{2,16}$(CNN) to $\text{U}_{2,16}$(CNN) and $\text{U}_{3,32}$(CNN), $\ell_1$DecNet+ $\text{U}_{3,32}$(Res18) to $\text{U}_{3,32}$(Res18) and $\text{U}_{3,32}$(Res34), $\ell_1$DecNet+ $\text{U}_{3,32}$(Effb0) to $\text{U}_{3,32}$(Effb0) and $\text{U}_{3,32}$(Effb2), $\ell_1$DecNet+ $\text{U}_{3,32}$(Mitb0) to $\text{U}_{3,32}$(Mitb0) and $\text{U}_{3,32}$(Mitb1), $\ell_1$DecNet+ $\text{MA}_{3,32}$(Res18) to $\text{MA}_{3,32}$(Res18) and $\text{MA}_{3,32}$(Res34), $\ell_1$DecNet+ $\text{Upp}_{3,32}$(Res18) to $\text{U}_{3,32}$(Res18) and $\text{Upp}_{3,32}$(Res34), respectively; see \reftable{tab:infer-3x} and \reftable{tab:train-3x}.
  
\reftable{tab:infer-3x} records the segmentation results in terms of AUC score on DRIVE, CHASE and CRACK datasets, along with the estimated number of learnable parameters of each network by PyTorch. Some segmetation results are shown in \reffigure{fig:seg} by zoom-in patches of the images and labels from DRIVE, CHASE and CRACK, along with the results from $\text{U}_{3,32}$(CNN), $\text{U}_{2,16}$(CNN) and $\ell_1$DecNet+$\text{U}_{2,16}$(CNN). \reftable{tab:train-3x} shows training, inference and storage costs in terms of time per backward-propagation step (time per bp step), number of multiply-accumulate operations in forward-propagation
(fwd MACs) and disk occupation, of the segmentation networks compared in \reftable{tab:infer-3x}. Therein, fwd MACs are measured by Python package DeepSpeed (0.11.1, cpu_py310) https://github.com/microsoft/DeepSpeed.

We can see that, in most cases, $\ell_1$DecNet+ architecture with a small-scale segmentation module performs the best, and achieves even better results than the enlarged version of the segmentation module, showing the effectiveness of the $\ell_1$DecNet sparse feature extractor and the flexibility of the choice of the segmentation module. We can also observe that $\ell_1$DecNet+ framework only introduce 0.02\% $\sim$ 1.42\% extra learnable parameters (2.5k) to the segmentation module. Overall speaking, the training costs (i.e. time per bp step) are comparable among the three models, $\ell_1$DecNet+ architecture, its segmentation module and enlarged version; while the inference cost (i.e. fwd MACs) and storage requirement (i.e. disk occupation) of our $\ell_1$DecNet+ architecture is lower than the  enlarged version of its segmentation module. These advantages arise from the fact that $\ell_1$DecNet can extract sparse features well and thus consistently help the subsequent segmentation procedures.

\begin{figure*}
  \centering
  \subfloat[DRIVE]
  {\includegraphics[width=0.33\textwidth]{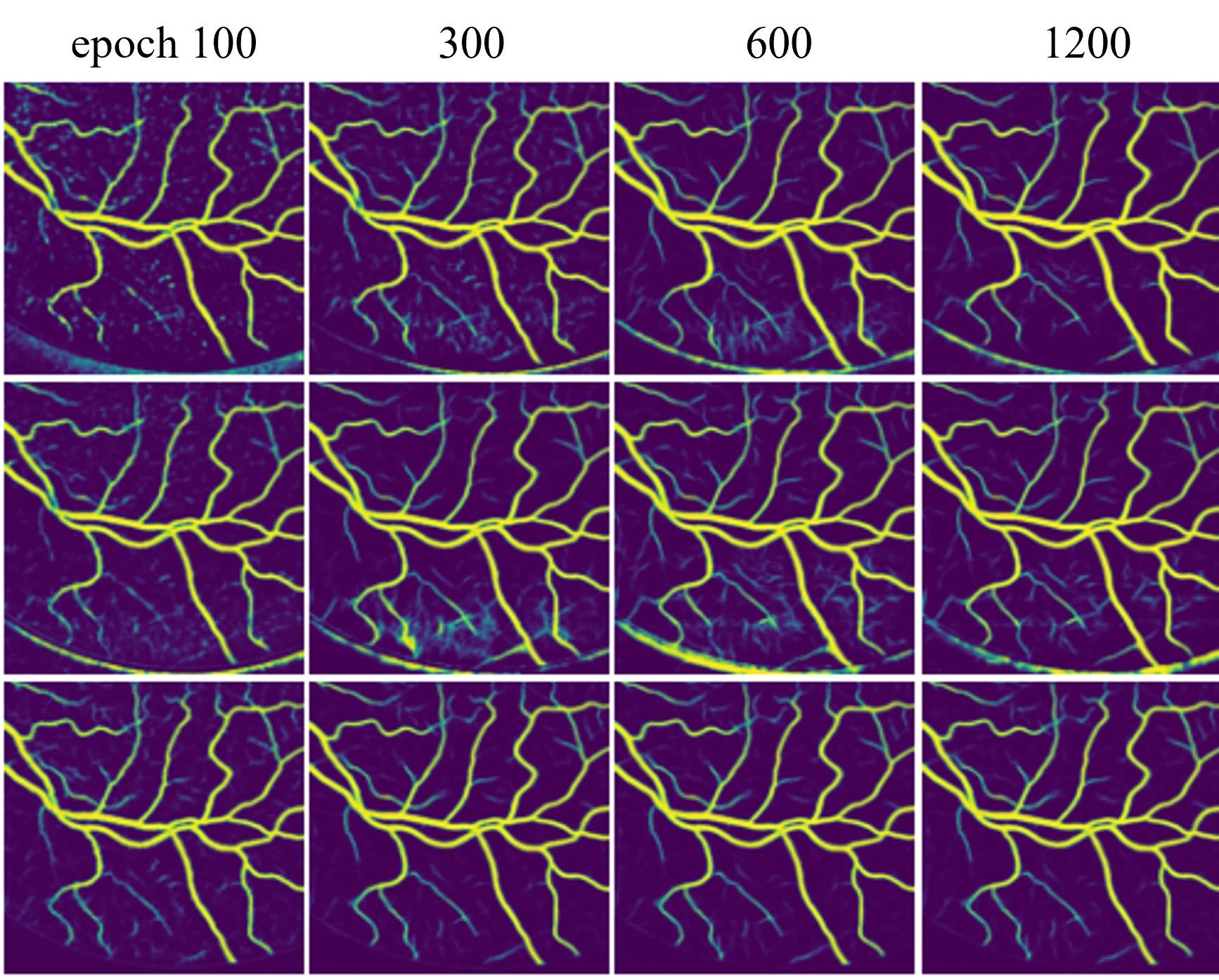}~}%
  \subfloat[CHASE]
  {\includegraphics[width=0.33\textwidth]{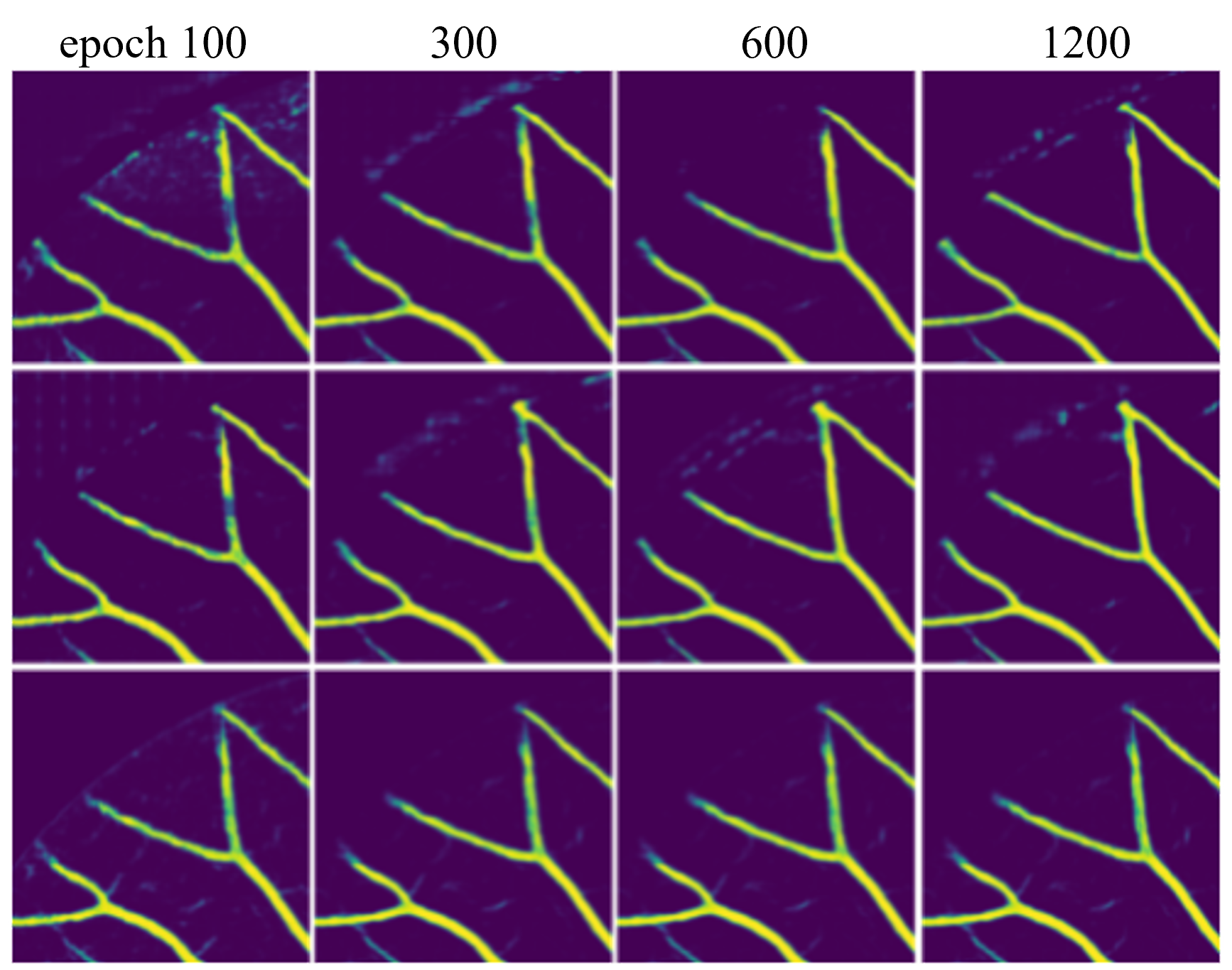}~}%
  \subfloat[CRACK]
  {\includegraphics[width=0.33\textwidth]{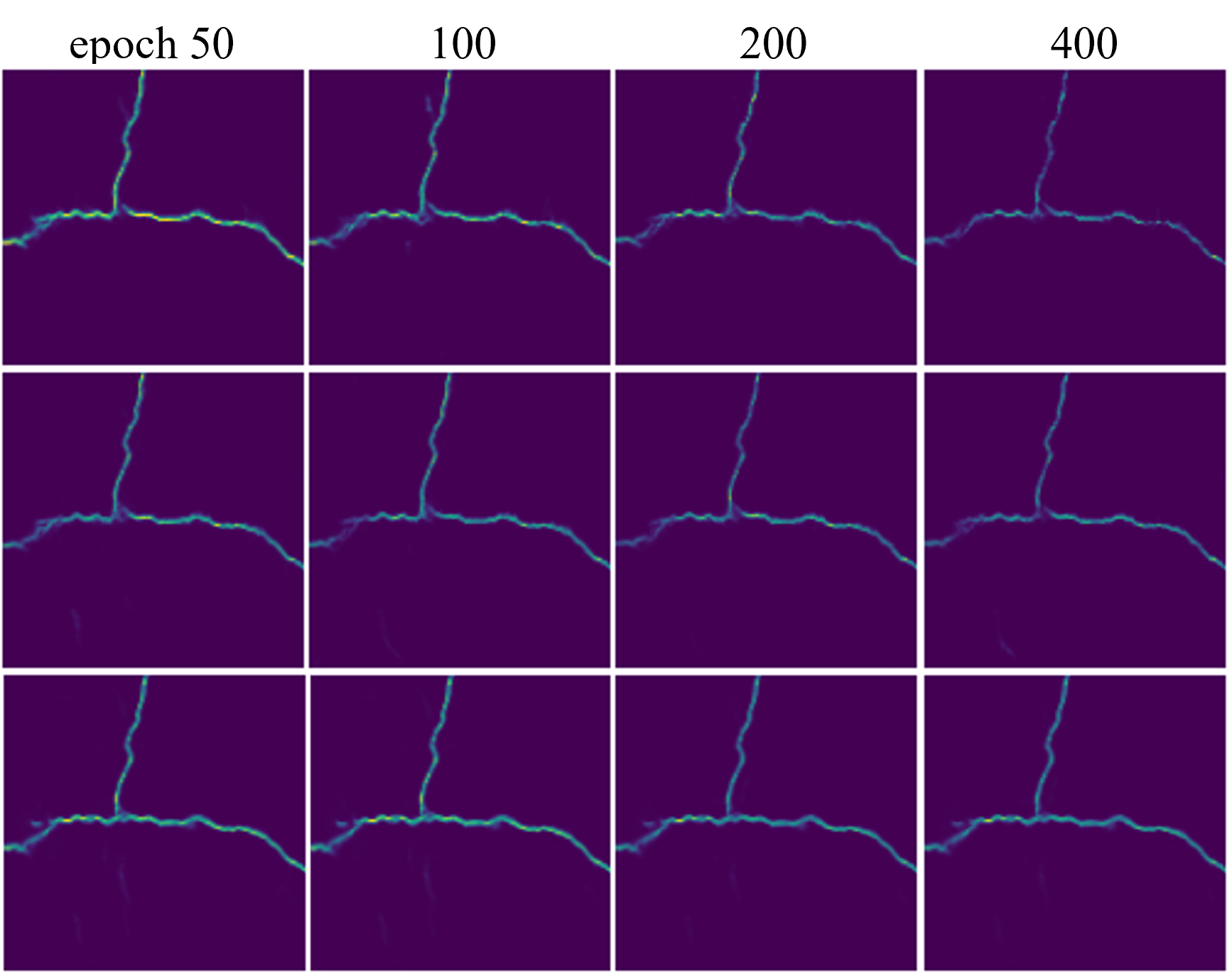}~}%
  \caption{Zoom-in of segmentation results on one test image by $\text{U}_{3,32}$(CNN) (row 1), $\text{U}_{2,16}$(CNN) (row 2) and $\ell_1$DecNet+$\text{U}_{2,16}$(CNN) (row 3) during training procedure on DRIVE (a), CHASE (b) and CRACK (c) datasets, respectively.} 
  \label{fig:train-seg-1}
\end{figure*}

\begin{figure*}
  \centering
  \subfloat[DRIVE]
  {\includegraphics[width=0.33\textwidth]{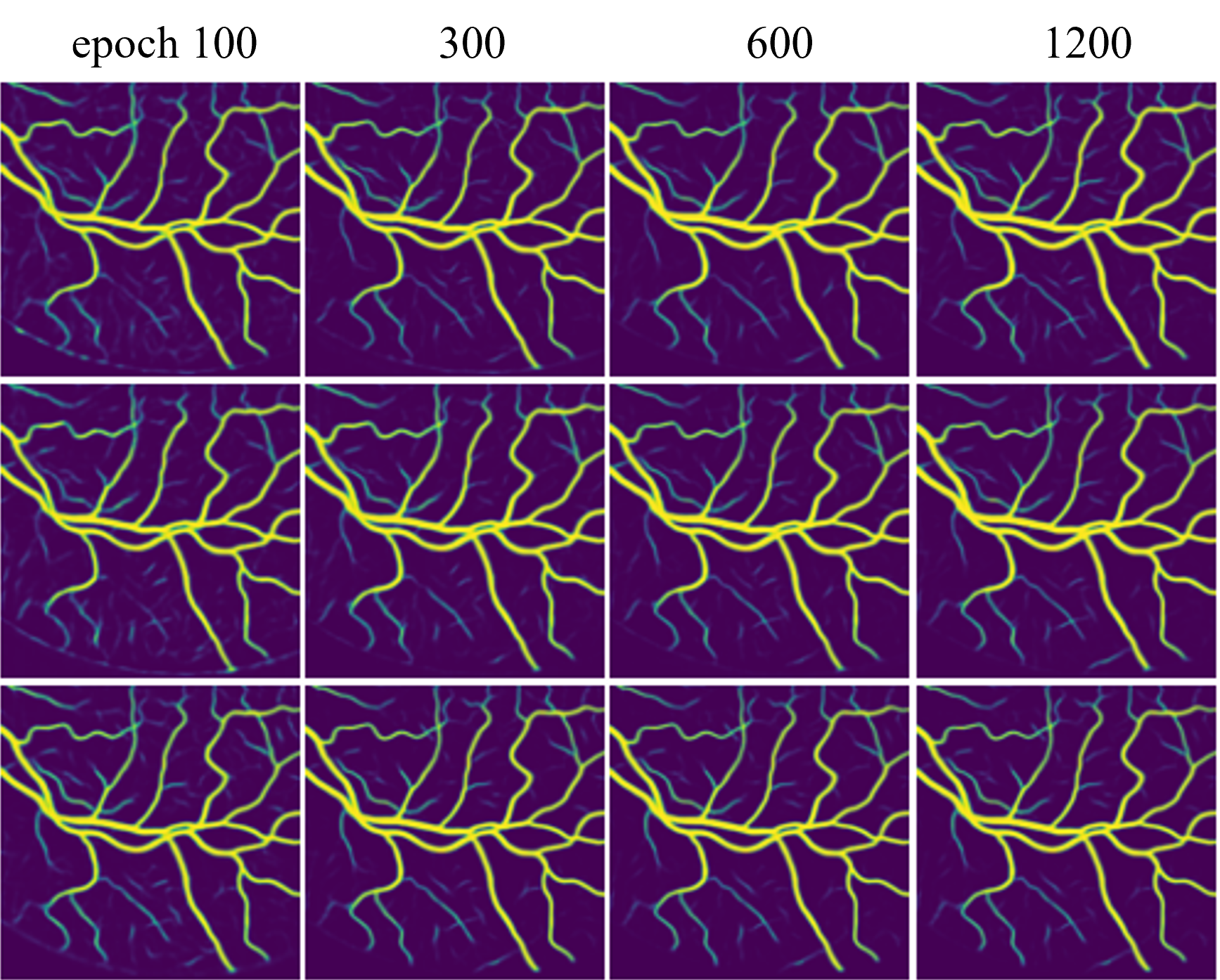}~}%
  \subfloat[CHASE]
  {\includegraphics[width=0.33\textwidth]{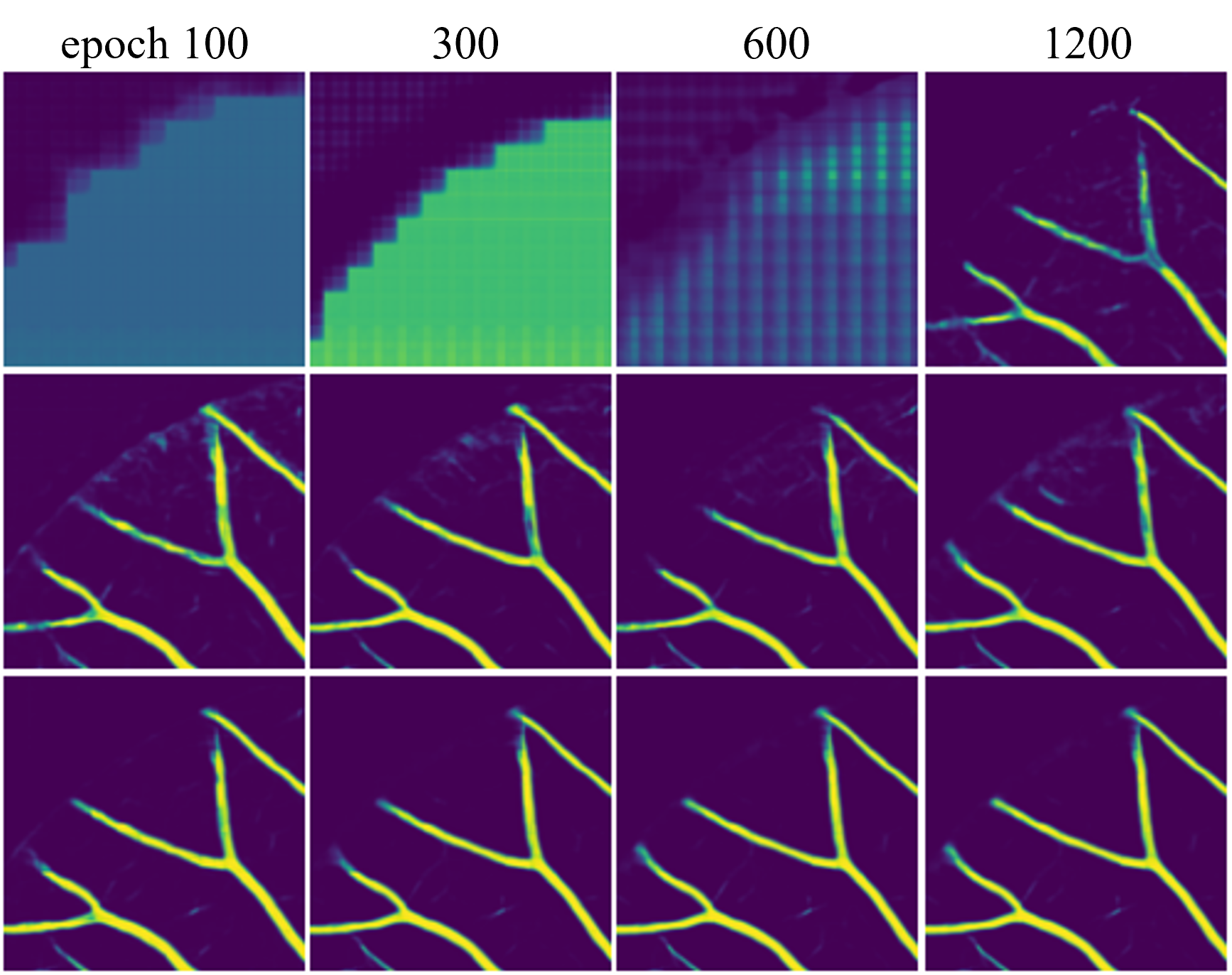}~}%
  \subfloat[CRACK]
  {\includegraphics[width=0.33\textwidth]{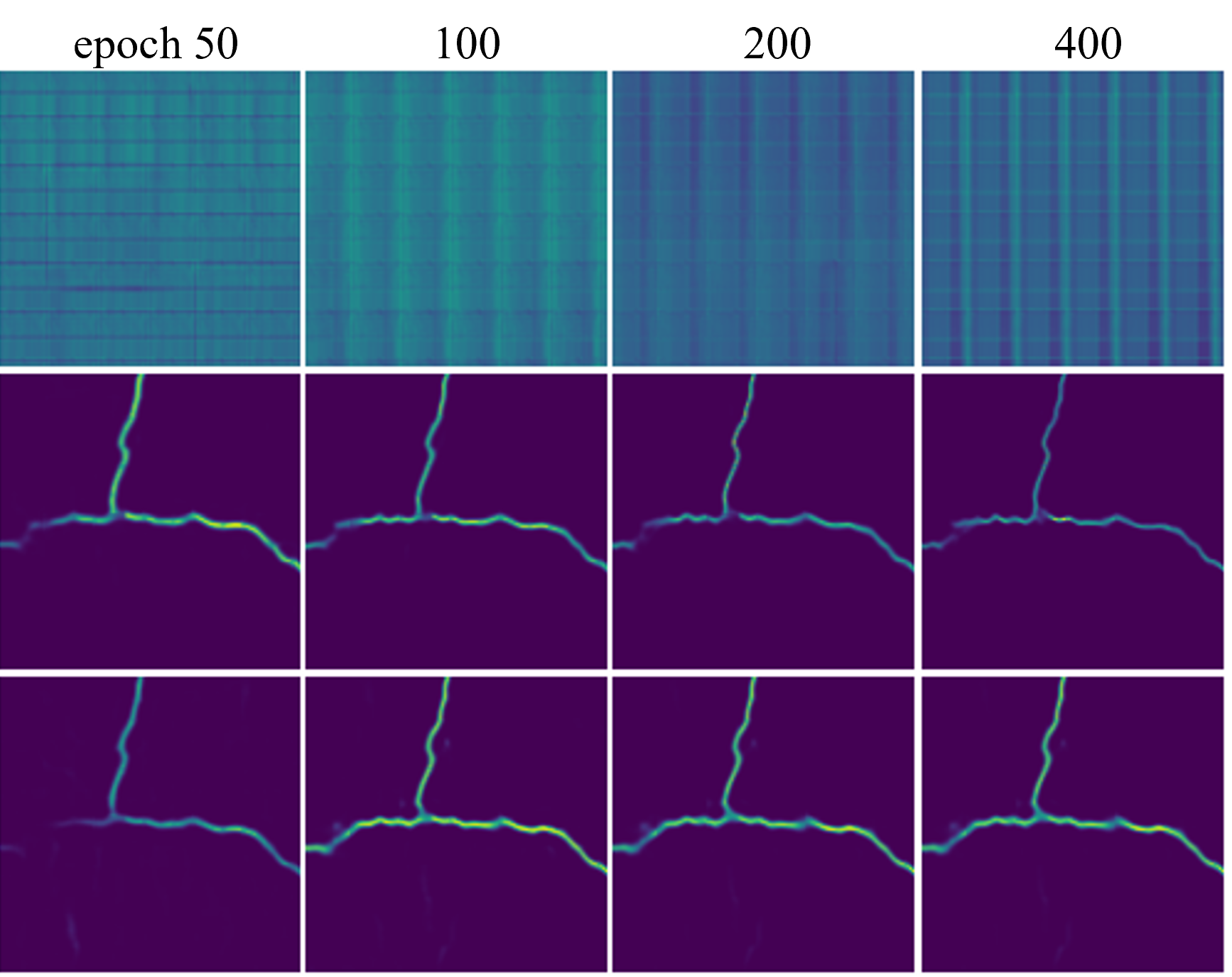}~}%
  \caption{Zoom-in of segmentation results on one test image by $\text{U}_{3,32}$(Mitb1) (row 1), $\text{U}_{2,16}$(Mitb0) (row 2) and $\ell_1$DecNet+$\text{U}_{2,16}$(Mitb0) (row 3) during training procedure on DRIVE (a), CHASE (b) and CRACK (c) datasets, respectively.} 
  \label{fig:train-seg-2}
\end{figure*}

\subsection{Comparison of the evolution of the segmentation inference during training procedure}

In this subsection, we compare the evolution of the segmentation inference during training procedure between our $\ell_1$DecNet+ architectures and other models. Such intermediate results are generated after training procedure, based on stored weights of networks at some specific epochs. We take the weights of networks at 100th, 300th, 600th and 1200th epochs of training on DRIVE and CHASE datasets, and 50th, 100th, 200th, 400th epochs of training on CRACK dataset. 

In \reffigure{fig:train-seg-1}, we show a comparison among $\text{U}_{3,32}$(CNN), $\text{U}_{2,16}$(CNN) and $\ell_1$DecNet+$\text{U}_{2,16}$(CNN) as a representation of the networks tested in \reftable{tab:infer-3x}. On DRIVE and CHASE datasets, \reffigure{fig:train-seg-1} shows that, $\text{U}_{3,32}$(CNN) and $\text{U}_{2,16}$(CNN) suffer an obvious illusion (the edge of the field of view) in the whole training procedure, while our $\ell_1$DecNet+$\text{U}_{2,16}$(CNN) keeps illusion-free; On CRACK dataset, it shows that the three models have similar behavior. \reffigure{fig:train-seg-2} illustrates a comparison among $\text{U}_{3,32}$(Mitb1), $\text{U}_{2,16}$(Mitb0) and $\ell_1$DecNet+$\text{U}_{2,16}$(Mitb0) as a special case that a simply enlarged model may collapse. It can be observed that the simply enlarged model $\text{U}_{3,32}$(Mitb1) takes more than 600 epochs to generate reasonable results on CHASE dataset, and even fails on CRACK dataset. Our $\ell_1$DecNet+ architecture with various segmentation modules performs well uniformly on three datasets. The reason is that our $\ell_1$DecNet+ architecture combines well the mathematical modeling and data-driven approaches.

\subsection{Validation of the sparse feature $v$ from $\ell_1$DecNet+}\label{sec:sparse-feature}
\begin{figure*}
  \centering
  \includegraphics[width=\textwidth]{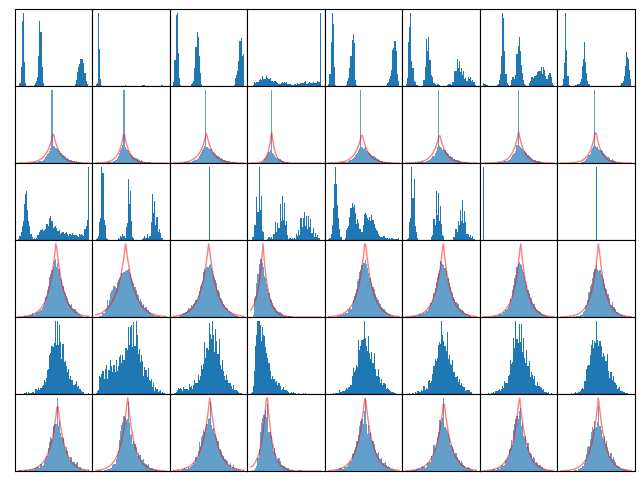}
  \caption{ Histograms in blue of the gray intensities of the random 8 patches of the first test image from DRIVE, CHASE and CRACK (row 1,3,5), and histograms of their feature $v$ by trained $\ell_1$DecNet+$\text{U}_{2,16}$(CNN) with Laplacian fitting curves in red (row 2,4,6). They are all normalized to the same height.}
  \label{fig:laplace-hist-v}
\end{figure*}

As we know, the entry values of a sparse vector obey a Laplacian distribution, which indeed induces the $\ell_1$ regularization in minimization models\cite{tibshirani_regression_1996}. In this subsection, we validate that the $v$ component of the $\ell_1$DecNet output does obey approximately a Laplacian distribution after network training. 

We use the experiment results of $\ell_1$DecNet+$\text{U}_{2,16}$(CNN) on DRIVE,CHASE and CRACK datasets. We compute the histograms of the gray intensities of some examples of input $f$, the histograms of those output feature $v$ and their Laplacian fitting curves. In \reffigure{fig:laplace-hist-v}, we take random 8 patches of the first image of each testing subset as the example for histogram illustration, where the distributions of patches of $f$ from DRIVE, CHASE and CRACK are quite different; see the first, third and fifth rows. The second, fourth and sixth rows show the histograms of their feature $v$ calculated by trained $\ell_1$DecNet+$\text{U}_{2,16}$(CNN) with Laplacian fitting curves by SciPy package. We can see that the Laplacian distribution prior of the feature $v$ is approximately kept through deep unfolding and network training, no matter what distribution $f$ follows.


\section{Conclusion and discussion}

In this paper, we proposed an unfolding network, namely $\ell_1$DecNet, from an $\ell_1$ regularized variational decomposition model and its scaled-ADMM solver. $\ell_1$DecNet outputs two components, one is a sparse feature characterized by an $\ell_1$ regularization, and the other is a dense feature (hard-to-mathematically-describe background) characterized by $\ell_1$ related regularization with some learnable sparsifying transformations. We also further constructed $\ell_1$DecNet+, a learnable architecture framework for sparse feature segmentation, by connecting $\ell_1$DecNet and some segmentation module. This architecture integrates well mathematical modeling and data-driven approaches. Benefited from the embedded sparsity prior in $\ell_1$DecNet, this architecture with any popular lightweight segmentation module can potentially achieve good performances stably in sparse feature segmentation problems. Experiments and comparisons on DRIVE, CHASE and CRACK datasets demonstrated such advantages. 

Because of the efficiency in learnable parameters and flexibility in module choice of our $\ell_1$DecNet+ framework, we can extend this network design strategy to more general applications. First, it is not difficult to extend our $\ell_1$DecNet+ framework to 3D case for volumetric data segmentation, by properly adapting the 2D convolutions to 3D. Second, our proposed $\ell_1$DecNet+ framework is for general sparse feature segmentation, and it is worthy to add application-specific constraints to it for concrete applications like topology and connectivity constraints for tubular structure segmentation problems.

\ifCLASSOPTIONcaptionsoff
  \newpage
\fi

\bibliographystyle{IEEEtran}
\bibliography{IEEEabrv,ref}

\end{document}